\documentclass[11pt,a4paper]{article}

\usepackage[dvipdfmx]{graphicx} 
\usepackage{cite}
\usepackage{bm}
\usepackage{amsmath}
\usepackage{amssymb} 
\usepackage{ulem}
\usepackage[legacycolonsymbols]{mathtools}
\usepackage{jheppub}
\newcommand{\ket}[1]{{\left\vert #1\right\rangle}}
\newcommand{\bra}[1]{{\left\langle #1\right\vert}}
\newcommand{\braket}[1]{\langle #1 \rangle}
\newcommand{\comm}[2]{{\left[#1, #2\right]}}
\newcommand{\expo}[1]{{\exp\left[ #1\right]}}
\newcommand{\abs}[1]{\left\vert #1 \right\vert}

\newcommand{\hintc}{\hat H_{\rm int}^{\rm (C)}}

\newcommand{\kk}{(\bm k,-\bm k)}

\newcommand{\ellef}{\ell_{\rm eff}}

\newcommand{\rhosq}{\rho_{\rm sq}}
\newcommand{\hth}{h_{\rm th}}
\newcommand{\hqg}{h_{\rm QG}}
\newcommand{\omegagw}{\omega_{\rm GW}}

\title{The Challenge of Detecting Quantum Nature of Gravitational Waves}

\author[a]{Yu Miyauchi,}
\emailAdd{miyauchi@tap.scphys.kyoto-u.ac.jp}
\affiliation[a]{Department of Physics, Kyoto University, Kyoto 606-8502, Japan}

\author[a]{Hidetoshi Omiya,}
\emailAdd{omiya@tap.scphys.kyoto-u.ac.jp}

\author[b]{Atsuhisa Ota,}
\emailAdd{aota@cqu.edu.cn}
\affiliation[b]{Department of Physics and Chongqing Key Laboratory for Strongly Coupled Physics, Chongqing University, Chongqing 401331, People's Republic of China}

\author[c,a]{Hiroki Takeda}
\emailAdd{takeda@tap.scphys.kyoto-u.ac.jp}
\affiliation[c]{The Hakubi Center for Advanced Research, Kyoto University, Kyoto 606-8501, Japan}

\author[a,d]{and Takahiro Tanaka}
\emailAdd{t.tanaka@tap.scphys.kyoto-u.ac.jp}
\affiliation[d]{
Center for Gravitational Physics and Quantum Information, Yukawa
Institute for Theoretical Physics, Kyoto University, Kyoto 606-8502, Japan
}

\date{}

\begin{document}
\abstract
{We investigate whether squeezing can provide an observable signature of quantum gravitational waves. Because a realistic detector couples only to a particular wave-packet mode, squeezing in global source modes need not remain observable. We show that inflationary two-mode squeezing reduces to an unsqueezed thermal state in the accessible one-mode sector, phase incoherence washes out squeezing in stochastic backgrounds, and the limited coverage of the solid angle of detectors strongly suppresses squeezing from isolated sources. We then show that source squeezing is not essential, {\it i.e.}, a quantized gravitational wave can generate a positive squeezing witness if the detector state is initially prepared in a squeezed state, whereas a classical external gravitational field cannot, producing only a displacement. However, the resulting signal is bounded by the extremely small graviton--detector coupling. Thus, detector squeezing can remove the need for squeezed incident waves, but not the suppression caused by weak gravitational interaction.
}

\maketitle

\section{Introduction}
\label{sec_intro}

Establishing whether gravity has quantum degrees of freedom
remains a central experimental challenge.  A decisive test must distinguish
the predictions of a quantum description of gravity from those of
an appropriately defined classical alternative, rather than merely detect a
small gravitational signal.  Gravitational waves provide a natural setting
for this question because they directly probe the radiative degrees of
freedom of gravity.  Evidence for their non-classical behavior would not by
itself constitute a test of a complete theory of quantum gravity, but it would
directly challenge an ordinary classical-field description of propagating
gravitational degrees of freedom.

Among the proposals to probe the quantum nature of gravitational waves, the search for squeezing is thought to offer a promising route. Direct detection of an individual graviton is generally considered impractical~\cite{Dyson:2013hbl}. A recent proposal instead considers a single quantum transition in a detector driven by a highly occupied gravitational wave~\cite{Tobar:2023ksi}. However, such a click does not establish that the incident field is quantum if its statistics can be reproduced by a classical stochastic field~\cite{Carney:2023nzz}. A stronger test must therefore probe statistics incompatible with a statistical ensemble of coherent waves~\cite{Glauber:1963tx,PhysRevLett.10.277}. Within Gaussian states,  squeezing provides a sharp sufficient witness of this kind. Note that the failure to observe squeezing in a specified mode does not establish that the gravitational wave is a classical field, whereas observed squeezing would provide evidence of nonclassical gravitational radiation. 

Gravitational wave squeezing has accordingly been investigated both as a state generation problem and as a measurement problem. On the generation side, inflationary vacuum fluctuations provide the canonical example. Cosmological expansion parametrically amplifies inflationary vacuum fluctuations, producing two-mode squeezing between modes with opposite momenta~\cite{Polarski:1995jg,Grishchuk:1998qz,Allen:1999xw}. Nonlinear effects in radiation from astrophysical sources have also been investigated as possible squeezing mechanisms~\cite{Kanno:2025how,Manikandan:2025dea,Guerreiro:2025mcu}, and more general frameworks for producing squeezed gravitational states have been proposed~\cite{Das:2025kyn}. On the measurement side, proposed probes of squeezing or its associated nonclassical statistics include interferometric measurements of quadrature fluctuations~\cite{Parikh:2020kfh,Kanno:2021gpt,Hertzberg:2021rbl}, Hanbury Brown--Twiss interferometry~\cite{Kanno:2018cuk,Kanno:2025fpz}, phonon counting~\cite{Manikandan:2025lfx,Toccacelo:2026hcz}, and graviton--photon conversion~\cite{Ikeda:2025uae}.

Generation and measurement studies do not, however, connect automatically, because squeezing is a property of a specified mode rather than a general property of the gravitational field. Source calculations usually characterize global plane-wave or spherical-wave modes. By contrast, a detector with finite spatial extent and observation time couples to a particular wave-packet mode $A$, selected by the detector response. The reduced state of $A$ is obtained by tracing out all orthogonal, inaccessible gravitational-wave modes. Tracing out the inaccessible modes can substantially alter the squeezing present in the source description. Moreover, an experiment does not read out the operator corresponding to the gravitational wave mode $A$ directly. The experiment instead measures the operator corresponding to a detector mode $b$, to which information about $A$ is transferred through the weak gravitational coupling. Source analyses often stop at the global source state, whereas detector analyses often begin by assuming the state of the incident gravitational wave mode. The resulting map from the source state, through the accessible mode $A$, to the detector state is therefore often left implicit. Whether squeezing at the level of the source survives both the reduction to $A$ and the subsequent transfer to $b$ is the central observability question addressed in this work.

Restricting the analysis to Gaussian states, we first examine how the reduction from the source state to the detector-accessible wave-packet mode affects squeezing. Inflationary gravitational waves provide a natural starting point, but their two-mode squeezed state reduces to an unsqueezed thermal one-mode state. We therefore next consider a hypothetical stochastic background composed of single-mode squeezed states. In this case, however, the squeezing phases of waves arriving from different directions cannot be coherently aligned, and phase averaging washes out the squeezing. This motivates the final case of an isolated source, for which the radiation effectively arrives from a single direction and avoids this phase-cancellation problem. Even then, the limited angular overlap between the global source mode and the local detector mode suppresses the observable squeezing by the squared ratio of the detector size to the source distance.

In the latter half of this paper, we ask whether the incident gravitational wave must be squeezed in order to distinguish a quantum field from a classical $c$-number external field.~\footnote{More general dynamical classical--quantum theories are also possible. Consistent nontrivial coupling between classical and quantum degrees of freedom generally involves stochastic dynamics, both in fundamentally hybrid formulations~\cite{Oppenheim:2018igd} and in effective descriptions emerging from environmental decoherence of an underlying fully quantum theory~\cite{Tomizuka:2026hmp}. We do not consider such dynamical classical--quantum theories here.}  If the detector is prepared in a controlled squeezed state, a quantized gravitational wave can generate a signature of squeezing in the transformed detector state even when the incident gravitational wave mode is not excited at all. By contrast, a classical external field only displaces the detector state and cannot generate such a signature. However, the quantum signature is bounded by a small coupling between the detector and the gravitational wave. Our optimistic estimates will show the coupling is extremely small throughout the frequency range considered.  Preparing a detector state in a controlled squeezed state can therefore remove the need for source squeezing, but it does not remove the suppression caused by weak gravitational coupling.

The plan of the paper is as follows.
In Sec.~\ref{sec_squeezing}, we briefly review Gaussian quantum states and introduce a witness that quantifies the effective degree of squeezing.
In Sec.~\ref{sec_linear response}, we formulate the linear interaction between gravitational waves and a local detector and identify the effective wave-packet mode that couples to the detector.
In Sec.~\ref{sec_vanishing}, we evaluate the witness for the effective mode in inflationary, stochastic-background, and isolated-source scenarios.
In Sec.~\ref{sec_epsilon}, we show that the witness for the detector mode is suppressed by the graviton--detector coupling and estimate its size.
Sec.~\ref{sec_conclusion} summarizes our results.
In this paper, we use natural units with $c=\hbar=1$.

\emph{Note added:}
During the preparation of this work, Ref.~\cite{Bao:2026prd} appeared on arXiv. The reference provides a general analysis of how nonclassical signatures of weakly coupled waves are suppressed at two distinct stages: through effective-mode coarse graining and through weak linear transfer to the detector, both of which overlap with the central logic of this work. Our work complements that general analysis with  a detailed analysis of gravitational-wave scenarios: the reduction of source squeezing upon projection onto detector-accessible modes for inflationary, stochastic-background, and isolated-source scenarios---including a geometric suppression bound---and estimates of the graviton--detector coupling based on representative gravitational-wave sensitivities across a broad frequency range.

\section{Squeezing and Witness} 
\label{sec_squeezing}

In this section, we review some standard properties of squeezing in quantum optics and introduce the witness $W$ that characterizes observable squeezing (see, e.g., Refs.~\cite{Agarwal:2013,Weedbrook:2011wxo}). 
Within the class of Gaussian states, the existence of measurable squeezing is equivalent to $W>0$. On the other hand, when $W\leq0$, the state can be represented as a statistical mixture of coherent states, and such a state is sometimes described as lacking quantumness. Although we will later discuss why this is not necessarily the case, in this section, rather than addressing quantumness itself, we summarize the basic properties of the witness as a criterion for determining whether a state can be identified as a squeezed state.

\subsection{Coherent States and Squeezed States}
We briefly review several well-known facts in the context of quantum optics. Restricting attention to one-mode Gaussian states, any pure state can be generated from the vacuum state by two operations: squeezing and displacement. Since displacement corresponds to shifting the origin in phase space, it is not directly related to squeezing when only pure states are considered. In general, however, under the condition that the density matrix is Gaussian, mixed states are also allowed. To represent a general Gaussian state, one must then also consider states constructed by a Gaussian ensemble of displaced squeezed states.

First, let $\hat a^\dag$ and $\hat a$ denote the creation and annihilation operators of a harmonic oscillator, respectively.
The coherent state is defined as an eigenstate of the annihilation operator $\hat a$ by
\begin{equation}
    \hat a\ket{\alpha}=\alpha \ket{\alpha}\,.
\end{equation}
This state is obtained by applying the displacement operator
\begin{equation}\label{eq:displace}
    \hat D(\alpha)=\expo{\alpha\hat a^\dagger-\alpha^* \hat a},
\end{equation}
to the vacuum state $\ket{0}$ as
\begin{equation}
    \ket{\alpha} = \hat D(\alpha)\ket{0}\,,
\end{equation}
which is immediately verified using the relation $\hat D^\dag(\alpha) \hat a \hat D(\alpha)=\hat a + \alpha$, derived using the Baker--Campbell--Hausdorff formula.\footnote{$\displaystyle \exp[-B] A \exp[B]=A+[A,B]+\frac1{2!}[[A,B],B]+\frac1{3!}[[[A,B],B],B] + \cdots$.} 

A single-mode squeezed vacuum state is defined as
\begin{align}
\label{eq_one_squeezed_state}
    |\xi\rangle
    =
    \hat S(\xi)\ket{0}\,,
\end{align}
where the squeezing operator is given by
\begin{align}
\label{eq_def_sqeezed_one}
    \hat S(\xi)=
    \exp\left[\frac 12\left(\xi \hat a^{\dagger2}- \xi^* \hat a ^2\right)\right]
    ,\quad
    \xi=re^{i\phi}\,.
\end{align}
The state $|\xi\rangle$ is the vacuum state annihilated by the annihilation operator
\begin{align}
    \hat{\tilde a}:=\cosh r\, \hat a -e^{i\phi}\sinh r\, \hat a^\dag\,,
\end{align} 
related to the original creation and annihilation operators by the Bogoliubov transformation.  
Namely, $\hat{\tilde a}\ket{\xi}=0$ holds.
This follows from the relation 
\begin{align}
  \hat S^\dag (\xi)\hat a \hat S(\xi)= \cosh r\,\hat a+ e^{i\phi} \sinh r\, \hat a^\dag\,. 
\end{align}

A Gaussian state, including a mixed state, is completely specified by the mean $\braket{\hat a}$ and the covariance matrix.
To define the covariance matrix, we first introduce $\delta \hat a \coloneqq \hat a-\braket{\hat a}$ and define the second-order correlations by
\begin{equation}
    \label{eq_def_correlation}
    N_a \coloneqq \braket{\delta\hat a^\dagger\delta \hat a}\,,
    \quad
    M_a \coloneqq \braket{\delta\hat a^2}\,.
\end{equation}
The correlator $N_a$ characterizes the fluctuation occupation, while $M_a$ determines
the quadrature-angle dependence of the fluctuations. 
Since $N_a$ is real and $M_a$ is complex, they specify the three independent real parameters of the covariance matrix.
Using the quadratures associated with $\hat a$,
\begin{equation}
\label{eq_quad_def}
     \hat X =\frac{\hat a+\hat a^\dagger}{\sqrt 2}\,,\quad  
     \hat P =\frac{\hat a-\hat a^\dagger}{\sqrt 2 i}\,,
\end{equation}
the $2\times2$ covariance matrix $V$ is defined by
\begin{align}
    V_{ij}
    \coloneqq\left\langle\frac12 \left(\Delta \hat R_i\Delta \hat R_j+\Delta  \hat R_j\Delta \hat R_i\right)\right\rangle\notag\,,
\end{align}
where $\Delta \hat R_i = \hat R_i-\braket{\hat R_i}$, with $\hat R_1=\hat X$ and $\hat R_2=\hat P$.
In terms of $M$ and $N$, the covariance matrix $V$ is expressed as
\begin{equation}
V=
    \begin{pmatrix}
        \frac12+N+\mathrm{Re}\,M & \mathrm{Im}\,M \\
        \mathrm{Im}\,M & \frac12+N-\mathrm{Re}\,M
    \end{pmatrix}\,,
\label{Eq:covariance_def}
\end{equation}
whose eigenvalues are
\footnote{Writing $M=\abs{M}e^{i\phi_M}$, the corresponding eigenvectors $\bm v_\pm$ are
\begin{equation}
\bm v_+=
    \begin{pmatrix}
        \cos \frac{\phi_M}{2}  \\
        \sin \frac{\phi_M}{2} 
    \end{pmatrix}
    ,\quad
    \bm v_-=
    \begin{pmatrix}
        -\sin  \frac{\phi_M}{2}  \\
        \cos \frac{\phi_M}{2} 
    \end{pmatrix}\,,
\end{equation}
respectively.}
\begin{equation}
    \lambda_\pm =\frac 12 + N \pm|M|\,.
\end{equation}
This matrix plays the role of the covariance of a probability distribution in phase space.
For a coherent state, $N=\abs{M}=0$, and the probability distribution is circularly symmetric.
The variance in every direction is $1/2$. This variance of a coherent state represents vacuum fluctuations.

For a squeezed state, variances in the major and minor semi-axes are $e^{2r}/2$ and $e^{-2r}/2$, respectively, with the principal axis tilted in the $\phi/2$ direction.
One finds that the distribution along the minor axis can be narrower than the vacuum fluctuations. Taking a classical statistical mixture can broaden a distribution but cannot narrow it.
Therefore, as a quantity characterizing quantum squeezing, we define the squeezing witness by
\begin{equation}
    W\coloneqq \frac12-\lambda_-=\abs{M}-N\,. 
\end{equation}
By construction, $W>0$ means that the minimum quadrature variance lies below the vacuum fluctuation level.
For a squeezed state, $N=\sinh^2 r$ and $M=e^{i\phi}\cosh r\sinh r$, and hence we have 
\begin{equation}
    W=\frac{1-e^{-2r}}{2}\,.
\end{equation}
Thus, $W$ is positive and approaches $1/2$ in the infinitely squeezed limit:
\begin{equation}
    0<W<\frac12\,.
\end{equation}

\subsection{Mixed States and Effective Squeezing}
A general one-mode Gaussian state, including mixed states, can be represented by a density matrix describing a Gaussian-weighted statistical mixture of displaced squeezed states (see, e.g., Refs.~\cite{Williamson:1936, Weedbrook:2011wxo}):
\begin{equation}
\label{eq_mixed_state}
    \rho = \int d^2\alpha G(\alpha) \hat D(\alpha)\rhosq\hat D^\dagger(\alpha)\,,\quad
   \rhosq = \hat S(\xi)\ket{0}\bra{0}\hat S^\dagger(\xi)\,.
\end{equation}
Here, $G(\alpha)$ is a Gaussian probability distribution used to form the statistical mixture.
For $r=0$, $\rho_{\rm sq}$ reduces to the vacuum state, and Eq.~\eqref{eq_mixed_state} becomes a statistical mixture of coherent states.
In the limiting case $G(\alpha)\rightarrow\delta^{(2)}(\alpha-\alpha_0)$, the state reduces to the pure state $\ket{\Psi}=\hat D(\alpha_0)\hat S(\xi)\ket{0}$.

The covariance matrix in Eq.~\eqref{eq_mixed_state} can be decomposed as
\begin{equation}
    V = V_{\rm sq}+V_{\alpha}\,.
\end{equation}
In other words, the total covariance matrix is simply the sum of the covariance matrix $V_{\rm sq}(\xi)$ of the underlying squeezed vacuum state, given in Eq.~\eqref{Eq:covariance_def}, and an additional contribution $V_{\alpha}$ arising from the classical statistical mixture over the displacement amplitudes described by $G(\alpha)$.
Here, we define the deviation of the complex displacement amplitude from its mean as $\delta\alpha:=\alpha-\mathbb{E}_G[\alpha]$, where the ensemble average with respect to $G(\alpha)$ is defined by
\begin{equation}
\mathbb{E}_G[f(\alpha,\alpha^*)] \coloneqq \int d^2\alpha\,G(\alpha)f(\alpha,\alpha^*)\,.
\end{equation}
Then, $V_\alpha$ is given by
\begin{align}
    V_\alpha
    &=
    \begin{pmatrix}
        N_\alpha+\mathrm{Re}\,M_\alpha & \mathrm{Im}\,M_\alpha \\
        \mathrm{Im}\,M_\alpha & N_\alpha-\mathrm{Re}\,M_\alpha
    \end{pmatrix}\,,
\end{align}
where $N_\alpha$ and $M_\alpha$ are defined as
\begin{align}
     N_\alpha
     &\coloneqq \mathbb{E}_G[\abs{\delta\alpha}^2] =\int d^2\alpha\, G(\alpha)\abs{\delta\alpha}^2,
     \\
     M_\alpha
      &\coloneqq \mathbb{E}_G[(\delta\alpha)^2]
     =\int d^2\alpha\, G(\alpha)(\delta \alpha)^2\,.
\end{align}
The Cauchy--Schwarz inequality implies
\begin{equation}
    \abs{M_{\alpha}} \leq N_{\alpha} \,.
    \label{Const_alpha}
\end{equation}
Therefore, by the triangle inequality and Eq.~\eqref{Const_alpha}, the following inequality must hold:
\begin{align}
    W=|M_{\rm sq} + M_{\alpha}|-N_{\rm sq}-N_{\alpha}
    &\leq |M_{\rm sq}|-N_{\rm sq}+|M_{\alpha}|-N_{\alpha} \notag
    \\
    &\leq |M_{\rm sq}|-N_{\rm sq} = W_{\rm sq}\,.
\end{align}

This inequality reflects the fact that the positive-semidefinite contribution $V_\alpha$ cannot reduce the variance along any quadrature from the underlying pure state.
It also implies that $W>0$ requires $W_{\rm sq}>0$. Since a statistical mixture of coherent states has $W_{\rm sq}=0$, a state with $W>0$ cannot be realized as such a mixture.
Conversely, when $W\leq0$, we have $\abs{M}\leq N$. Therefore, one can choose $r=0$, $N_\alpha=N$ and $M_\alpha =M$ such that satisfies Eq.~\eqref{Const_alpha}. Thus, we find that the state can be represented as a statistical mixture of coherent states.
To conclude, within the class of Gaussian states, $W>0$ is a necessary and sufficient condition for squeezing to be indispensable in representing the state. 
We therefore focus on $W$ in the following discussion.

\section{Linear Response and Effective Mode}
\label{sec_linear response}

The physical quantities of interest must, of course, be observable. This requires us to consider a detector, although the detailed detector setup is irrelevant to the following discussion. It is sufficient to impose the following simple assumptions. First, we assume that the gravitational-wave signal is weak and interacts with the detector linearly in the gravitational-wave operator. This assumption entails no loss of generality for detecting gravitational waves. 

We further assume that the detector quantum state can be described using a linear operator of a single oscillator, with its annihilation operator denoted by $\hat{b}$. One could consider models in which detector operators interact nonlinearly, but because the coupling between gravitational waves and matter fields is already weak, there appears to be no advantage in deliberately suppressing the linear response and considering nonlinear interactions. One could also consider linear combinations of oscillators with different frequencies, but this would only enhance the loss of quantum coherence. This assumption is therefore also sufficiently general.

For quantum gravitational waves, we expand the field as
    \begin{equation}
        \hat h_{ij}(t,\bm x)=\sum_\lambda \int \frac{d^3\bm k}{(2\pi)^3} e^\lambda_{ij}(\bm k)\left[u_k(t)\hat a _\lambda(\bm{k})e^{i\bm k\cdot \bm x}+({\rm h.c.})\right],
    \end{equation}
using appropriately normalized plane waves
    \begin{equation}
        u_k(t)=\frac{2}{M_\mathrm{pl}}\frac{e^{-i\omega_k t}}{\sqrt{2\omega_k}}\,,
    \label{Eq:mode_function}
    \end{equation}
and polarization tensors $e^\lambda_{ij}$. Latin indices $i,j,\ldots$ run over the three spatial dimensions $1,2,3$. For each wave vector ${\bm k}$ and polarization mode $\lambda$, we introduce creation and annihilation operators $\hat a^\dag_\lambda(\bm{k})$ and $\hat a_\lambda(\bm{k})$ and impose the commutation relation
\begin{align}
    [\hat a_\lambda({\bm k}),\hat a_{\lambda'}^\dagger({\bm k'})]
       =(2\pi)^3\delta_{\lambda \lambda'}\delta^{(3)}(\bm k-\bm k^\prime)\,.
\end{align}
The normalization in Eq.~\eqref{Eq:mode_function} explicitly shows that the amplitude of the gravitational wave mode is Planck-suppressed.

In general, the interaction between a gravitational wave and the matter field constituting the detector can be written in terms of its energy--momentum tensor $\hat T_{\mu\nu}$ as
    \begin{align}
        \hat H_\mathrm{int}(t)
        &=-\frac 12\int_V d^3\bm x \,\hat h_{ij}(t,\bm x)\hat T^{ij}(t,\bm x)\notag&
        \\
        &=-\frac 12\sum_\lambda\int \frac{d^3\bm k}{(2\pi)^3}\left[u_k(t)\hat a _\lambda(\bm{k})\hat {\mathcal T}_\lambda(t,\bm k)+({\rm h.c.})\right]\, ,
\label{eq:quantum_model}
    \end{align}
where
  \begin{align}
         \hat {\mathcal T}_\lambda(t,\bm k)
         &=\int_Vd^3\bm x \, e^{i\bm k\cdot \bm x} e^\lambda_{ij}(\bm k) \hat T^{ij}(t,\bm x)~.
 \end{align}
The operator $\hat {\mathcal T}$ is the transverse--traceless stress component to which the plane-wave gravitational-wave mode labeled by $(\lambda,\bm k)$ couples.  After linearizing the detector response, we retain the part involving the relevant detector mode $\hat{b}$,
\begin{align}\label{eq:responce_linear}
    \hat {\mathcal T}_\lambda(t,\bm k)
    \simeq
    \Gamma_{\lambda}(\bm k)\, \hat b \, e^{-i\omega_b t}
    +
    \Lambda_\lambda(\bm k)\, \hat b^\dagger \, e^{i\omega_b t}\,.
\end{align}
Here, $\omega_b$ is the frequency of the detector mode.  The coefficients $\Gamma_\lambda$ and $\Lambda_\lambda$ encode the spatial profile and internal dynamics of the particular detector. 

We restrict the evolution to a finite observation interval $T$. Let $\chi(t)$ be a window function with support of duration $T$. The time evolution is governed by the unitary operator
\begin{align}
   \hat{U} 
    &=
    \mathsf{T}
    \exp\left[
        -i\int dt\,\chi(t)\hat H_{\rm int}(t)
    \right]\, ,
\end{align}
where $\mathsf{T}$ denotes time ordering.
To leading order in the weak
coupling, we have
\begin{align}
    \hat{U}
    \sim
    1 - i \int dt\,\chi(t)\hat H_{\rm int}(t)~.
\end{align}
Substitution of the linearized detector response~\eqref{eq:responce_linear} produces terms oscillating at the sum $\omega_k+\omega_b$ and difference $\omega_k-\omega_b$ of the frequencies. The former is associated with $\hat a_\lambda\hat b$, whereas the latter is associated with $\hat a_\lambda\hat b^\dagger$, which describes the conversion of the initially excited gravitational wave mode to detector mode.

We focus on wave packets with $\omega_k\simeq\omega_b$.  For an observation duration $T\gg\omega_b^{-1}$, the oscillating part with frequency $\omega_k + \omega_b \sim 2\omega_b$ vanishes after time integration. The rotating-wave approximation then gives
\begin{align}
    -i\int dt\,\chi(t)\hat H_{\rm int}(t)
    \sim
    \sum_\lambda
    \int\frac{d^3\bm k}{(2\pi)^3}
    \left[
        f_\lambda(\bm k)
        \hat a_\lambda(\bm k)\hat b^\dagger
        -
        f_\lambda^*(\bm k)
        \hat a_\lambda^\dagger(\bm k)\hat b
    \right],
\end{align}
where
\begin{align}
    f_\lambda(\bm k)
    &=
    \frac{i}{M_{\rm pl}}
    \int dt\,\chi(t)
    \frac{\Lambda_\lambda(\bm k)}{\sqrt{2\omega_k}}\,
    e^{-i(\omega_k-\omega_b)t}\,.
\end{align}
The response function $f_{\lambda}(\bm{k})$ contains the frequency, polarization, spatial, and temporal selectivity of the detector.

Now we introduce the detector-accessible gravitational-wave mode as 
\begin{align}\label{eq:reduced_A}
    \hat A
    &=
    \sum_\lambda
    \int\frac{d^3\bm k}{(2\pi)^3}
    c_\lambda(\bm k)\hat a_\lambda(\bm k)~,
\end{align}
where
\begin{align}
    c_\lambda(\bm k)
    &=
    \frac{f_\lambda(\bm k)}{\epsilon}~,
    &
    \epsilon^2
    &=
    \sum_\lambda
    \int\frac{d^3\bm k}{(2\pi)^3}
    \abs{f_\lambda(\bm k)}^2~.
\end{align}
The coefficient $c_\lambda(\bm k)$ specifies the normalized wave packet selected by the detector, whereas the dimensionless parameter $\epsilon$ specifies the strength with which the gravitational wave mode is transferred to the detector readout mode. Note that the commutation relations of the plane-wave operators imply $[\hat A,\hat A^\dagger]=1$. With the definitions of $\hat A$ and $\epsilon$ above, the microscopic unitary evolution can be effectively represented as
\begin{align}\label{Eq:general_interaction}
	\hat U_{\rm eff}=\exp\left[\epsilon\left(\hat A\hat b^\dagger-\hat A^\dagger\hat b\right)\right]~,
\end{align}
whose expansion agrees with the microscopic evolution at leading order. For the following analysis, we utilize this effective time evolution operator.

Since the detector has access only to the effective mode $\hat{A}$, the detector output can depend on the initial gravitational-wave density operator $\rho_{\rm GW}$ only through the reduced state
\begin{equation}
    \rho_A
    =
    \operatorname{Tr}_{A^\perp}\rho_{\rm GW}\,,
\end{equation}
where $A^\perp$ denotes all gravitational-wave modes orthogonal to the accessible mode $\hat A$. Therefore, the information about the initial gravitational-wave modes is inevitably reduced. To express this reduction independently of the choice of source basis, let $\alpha$ denote a collective mode label, such as $(\lambda,\bm k)$ for plane waves or $(\ell,m)$ for spherical waves.  We may then write
\begin{align}
    \hat A &= \sum_\alpha c_\alpha\hat a_\alpha,
    &
    \sum_\alpha\abs{c_\alpha}^2 &= 1~,
    \label{eq:accecble_mode}
\end{align}
as in Eq.~\eqref{eq:reduced_A}.
The second-order correlations of the source are given by 
\begin{align}
    N_{\alpha\beta}
    &=
    \braket{\delta\hat a_\alpha^\dagger
        \delta\hat a_\beta}~,
    &
    M_{\alpha\beta}
    &=
    \braket{\delta\hat a_\alpha \delta\hat a_\beta}~,
\end{align}
where $\delta\hat a_\alpha=\hat a_\alpha-\braket{\hat a_\alpha}$.
The corresponding correlations of the accessible mode $\hat A$ are
\begin{align}
    N_A
    &=
    \braket{\delta\hat A^\dagger\delta\hat A}
    =
    \sum_{\alpha,\beta}
    c_\alpha^*c_\beta N_{\alpha\beta}\,,
    \\
    M_A
    &=
    \braket{(\delta\hat A)^2}
    =
    \sum_{\alpha,\beta}
    c_\alpha c_\beta M_{\alpha\beta}\,.
    \label{eq:accecible_N_M}
\end{align}
The squeezing witness of the detector-accessible gravitational-wave mode is now given by
\begin{align}
    W_A = \abs{M_A}-N_A~.
\end{align}
For a one-mode Gaussian state, $W_A>0$ is equivalent to a quadrature variance below the vacuum level.  If  $W_A\leq0$, the reduced Gaussian state $\rho_A$ admits a representation as a statistical mixture of coherent states. 

Through the effective unitary evolution in Eq.~\eqref{Eq:general_interaction}, the detector probes the original gravitational-wave state only via the effective mode defined in Eq.~\eqref{eq:reduced_A}. Therefore, any nonclassicality of the original state must survive the projection onto this mode in order to be accessible to the detector. We thus use $W_A$ to characterize the squeezing retained in the detector-accessible mode. In the next section, we evaluate the extent to which the witness is reduced by the above projection for several gravitational-wave sources.

\section{Reduction of Squeezing in Effective Modes}
\label{sec_vanishing}
In this section, we show that in any realistic scenario, the effective mode coupled to the detector explained in the previous section cannot be a squeezed state. In what follows, for simplicity, we consider gravitational-wave modes generated at the source as squeezed-vacuum states, neglecting the possibility of displacement, and hence $\delta\hat a=\hat a$.

\subsection{Two-Mode Squeezed State of Inflationary Gravitational Waves}
\label{Subsec_inflation}
The first candidate that comes to mind for a source of squeezed gravitational waves is inflation~\cite{Polarski:1995jg,Grishchuk:1998qz,Allen:1999xw}. We will see, however, that the observed inflationary gravitational-wave state does not exhibit squeezing at all.

The squeezing generated during inflation is usually characterized in terms of plane-wave modes. Consider the positive-frequency mode function $\tilde u_{\bm k}$ that defines the initial vacuum state. During inflation, once the mode becomes superhorizon, its evolution is dominated by the growing mode, and the mode function $\tilde u_{\bm k}$ can be chosen to be approximately real. After inflation, we choose the natural positive-frequency mode function $u_{\bm k}(t)$ such that it is approximately proportional to $e^{-i\omega t}$. In terms of this natural positive-frequency mode and its negative-frequency counterpart, the mode function $\tilde u_{\bm k}$ evolved through the super-horizon regime can be approximately expressed as $\tilde u_{\bm k}\propto u_{\bm k}+u^*_{-\bm k}$. Since the spatial dependence of $\tilde u_{\bm k}$ is proportional to $e^{i{\bm k}\cdot{\bm x}}$, the negative-frequency contribution with the same spatial dependence is $u^*_{-\bm k}$, rather than $u^*_{\bm k}$. This shows that the state is a two-mode squeezed state with correlations between the $\kk$ modes. More precisely, $\tilde u_{\bm k}$ can be written as
\begin{align}
  \tilde u_{\bm k}= \cosh r_{\bm k}u_{\bm k}+e^{-i\phi_{\bm k}} \sinh r_{\bm k}u^*_{-\bm k}\,,
  \label{Eq:inflation_bogo}
\end{align}
where $r_{\bm k}$ and $\phi_{\bm k}$ are the squeezing parameter and squeezing phase, respectively.

As discussed above, $\tilde u_{\bm k}$ becomes approximately real up to an overall convention-dependent phase. If the phase of $u_{\bm k}$ is chosen such that it is real before horizon re-entry, this property implies that the positive- and negative-frequency components are approximately in phase and have approximately equal amplitudes $u_{\bm k}\sim u^*_{-\bm k}$, requiring $\phi_{\bm k}\approx0$ and $r_{\bm k}\gg1$. In particular, the latter condition indicates that the state is highly squeezed. 

Corresponding to the mode-function relation in Eq.~\eqref{Eq:inflation_bogo}, the annihilation operator $\hat a_{\bm k}$ associated with the natural post-inflationary vacuum is related to the operator $\hat{\tilde a}_{\bm k}$ associated with the inflationary vacuum through the Bogoliubov transformation:
\begin{align}
  \hat a_{\bm k}=\cosh r_{\bm k} \hat{\tilde a}_{\bm k} 
     + e^{
     i\phi_{\bm k}}\sinh r_{\bm k} \hat{\tilde a}_{-\bm k}^\dag\,.
\end{align}
 We define the two-mode squeezing operator by
\begin{align}
    \hat S_2(\xi)
    =\prod_k\exp\left[
        \xi_{\bm k} \hat a_{\bm k}^\dagger\hat a_{-\bm k}^\dagger-
        \xi^*_{\bm k} \hat a_{\bm k} \hat a_{-\bm k}
    \right]\,, \quad
    \xi_{\bm k}=e^{i\phi_{\bm k}} r_{\bm k}\,.
\end{align}
Therefore, the vacuum state naturally prepared during inflation is given by
\begin{align}
    |\tilde 0\rangle
    =
    \hat S_2(\xi)|0\rangle \,.
    \label{Eq:state_inflation}
\end{align}
This state satisfies $\hat{\tilde a}_{\bm k}|\tilde 0\rangle=0$, which follows immediately from the relation
\begin{align}
    \hat S_2^\dag(\xi)\, \hat a_{\bm k}\, \hat S_2(\xi)
    = \cosh r_{\bm k}\,\hat a_{\bm k} +
    e^{i\phi_{\bm k}}\sinh r_{\bm k}\,\hat a_{-\bm k}^\dag\,.
 \label{eq:basic_twomode}
\end{align}

Next, we consider the observation of inflationary gravitational waves. The observed gravitational wave mode corresponds to the gravitational waves that pass through the detector during the observation time. We therefore need to consider the corresponding wave-packet modes rather than plane-wave modes. Let us consider a wave packet centered at $\bm x=\bm y_i$ in position space and at $\bm k=\bm p_j$ in momentum space at horizon re-entry. Here, $(i,j)$ are discrete labels specifying the position and wave-vector centers, respectively, and we choose the labeling such that $\bm p_{-j}=-\bm p_{j}$. Such a wave packet can be constructed as an appropriate linear combination of plane waves:
\begin{align}
    \tilde v_{i,j}(x)=\int \frac{d^3k}{(2\pi)^3}C_{i,j}({\bm k})\tilde u_{\bm k}
    &=\int \frac{d^3k}{(2\pi)^3}C_{i,j}({\bm k})\left(\cosh r_{\bm k}u_{\bm k}+e^{-i\phi_{\bm k}}\sinh r_{\bm k}u^*_{-\bm k}\right)\cr
    &\simeq \cosh r_{\bm p_j} v_{i,j}(x) + e^{-i\phi_{\bm p_j}}\sinh r_{\bm p_j} v^*_{i,-j}(x)\,,
    \label{wavepacket}
\end{align}
where $\xi_{\bm k}$ has been approximated as constant within each wave packet, $\xi_{\bm p_j}$. For this approximation to be valid, we consider that the spatial extension of the wave packets is sufficiently large, and hence only a tiny range of the momentum modes contributes to constructing the wave packets. We emphasize that this condition is not an assumption, but simply a choice of parameters that are freely adjustable in our setup.
Here, we define 
\begin{align}
    &v_{i,j}(x) \coloneqq \int\frac{d^3\bm k}{(2\pi)^3}\,                  C_{i,j}(\bm k) u_{\bm k}(x)\, ,
    \cr
    &v^*_{i,-j}(x) \coloneqq \int \frac{d^3\bm k}{(2\pi)^3}\, C_{i,j}(\bm k) u_{-\bm k}^*(x)\,.\label{Eq_vij}
\end{align}
The coefficients $C_{i,j}(\bm k)$ are peaked around $\bm k=\bm p_j$ and contain a position-dependent phase, schematically given by
$C_{i,j}(\bm k)\propto e^{-i\bm k\cdot\bm y_i}g(\bm k-\bm p_j)$. We choose them to form a complete orthonormal set satisfying
\begin{equation}
    \int \frac{d^3k}{(2\pi)^3}C_{i,j}({\bm k}) C^*_{i',j'}({\bm k})=\delta_{ii'}\delta_{jj'}\,.
  \label{eq:orthonormal}
\end{equation}
Denoting the annihilation operator associated with $v_{i,j}(x)$ by $\hat a_{i,j}$, the two-mode squeezed state in Eq.~\eqref{Eq:state_inflation} is expressed in terms of the wave-packet modes as
\begin{equation}
    \ket{\tilde 0} = \expo{\int\frac{d^3 k}{(2\pi)^3}\xi_{\bm k}\hat a^\dagger_{\bm k}\hat a^\dagger_{-\bm k}- ({\rm h.c.})}\ket{0}
    \approx\expo{\sum_{i,j}\xi_{{\bm p}_j}\hat a^\dagger_{i,j}\hat a^\dagger_{i,-j}-{\rm ({\rm h.c.})}}\ket{0}\,, 
\end{equation}
where the second equality follows from Eq.~\eqref{eq:orthonormal}, the relation $C_{i,j}({\bm k})=C_{i,-j}^*(-{\bm k})$, which follows from \eqref{Eq_vij}, and the unitary transformation relation $\displaystyle {\hat a_{i,j}^\dag}=\int\frac{d^3 k}{(2\pi)^3}C_{i,j}({\bm k}){\hat a_{\bm k}^\dag}$. The second equality is approximate because $\xi_{\bm k}$ is replaced with the representative value $\xi_{{\bm p}_j}$. 

Recall that at horizon re-entry, $u_{\bm k}$ is approximately real and $\phi_{\bm k}\simeq0$. The two wave packets $v_{i,j}$ and $v^*_{i,-j}$ are therefore both localized around $\bm x=\bm y_i$. It would be quite natural to find that mutually correlated pairs, created  during inflation so as to satisfy momentum conservation, are both in the same horizon patch at the horizon re-entry once wave packets are formed. Their subsequent propagation directions can be seen directly from Eq.~\eqref{Eq_vij}. In the sub-horizon regime, their centers approximately evolve as
\begin{equation}
\bm x_{i,\pm j}(\eta) \simeq \bm y_i \pm \hat{\bm p}_j(\eta-\eta_*)\,,
\end{equation}
where the plus sign corresponds to $v_{i,j}$, while the minus sign corresponds to $v^*_{i,-j}$. Here, $\eta$ denotes the conformal time, $\eta_*$ specifies the time at horizon re-entry, and $\hat{\bm p}_j=\bm p_j/|\bm p_j|$. Thus, the two wave packets propagate in opposite directions after horizon re-entry.

To be precise, we consider a general normalized detector-accessible mode originating from a patch labeled by $i$, 
\begin{align}
    \hat A
    =
    \sum_{j>0}c_{j}\hat a_{i,j}\,,
    \qquad
    \sum_{j>0}|c_{j}|^2=1\,,
    \label{eq:detector_wavepacket_superposition}
\end{align}
where $c_{j}$ specifies the accessible wave packet, and  
$j > 0$ labels the modes that intersect the detector during the observation.

Then the second moments of $\hat A$ are given by
\begin{align}
\label{eq_def_2mode_corr}
    & N_A=\langle \hat A^\dagger \hat A\rangle
       =\sum_{j>0} |c_j|^2 \sinh^2 r_{j} \,,
    \\
    & M_A=\langle \hat A^2\rangle
             = 0\,, 
\end{align}
directly from 
\begin{align}
    \hat S_2^{\dag}(\xi_j)\, \hat a_{i,j}\, \hat S_2(\xi_j)=\cosh r_{j}\, \hat a_{i,j}+e^{i\phi_{j}}\sinh r_{j}\,\hat a_{i,-j}^\dag\,, 
\end{align}
which is analogous to Eq.~\eqref{eq:basic_twomode}.
Therefore, the witness for the effective mode is
\begin{equation}
    W_A= -N_A\,.
\end{equation}
Since $r_j\gg1$, $W_A$ is very large and negative.

This is analogous to Hawking radiation, for which the reduced state accessible to an observer is well known to be thermal. In both cases, the global state is a two-mode squeezed state, while one member of each correlated pair is unobservable. The observable mode is therefore in a thermal mixed state, resulting in a large negative value of the witness. The claim given in this subsection is very close to what was discussed in Ref.~\cite{Allen:1999xw}. Our analysis makes clear that the squeezing signature cannot be seen in the inflationary primordial gravitational waves, however long the observation time would be extended in the future direction.

\subsection{Background Gravitational Waves with Single-Mode Squeezing}
In the previous subsection, we found that even the very large and most plausible inflationary squeezing is not observable, because it is two-mode squeezing. In this subsection, although the possibility is much more remote, we examine whether squeezing of a stochastic gravitational-wave background could be observed, assuming a hypothetical mechanism that generates single-mode squeezing in each wave packet.
We consider a quantum state given by
\begin{align}
\label{eq_washout_state}
    |\Psi\rangle = \prod_j \expo{\frac 12 \left(\xi_j\hat a^{\dagger2}_j-\xi_j^* \hat a_j^2\right)}\ket{0}\,,
\end{align}
where $j$ is an index labeling patches covering the entire sky seen by the local detector ($\bm x_D=0$).
The operator $\hat a^{\dagger}_j$ creates a gravitational-wave mode arriving from the $j$th patch, and $\xi_j$ characterizes the squeezing of that mode. We assume that the angular extent of each patch is determined by the spatial scale over which gravitational-wave coherence is maintained, so that $\Delta \Omega\sim ($patch size$/$propagation distance$)^2$.~\footnote{Suppose, for the sake of argument, that the squeezing phases $\phi_j$ of gravitational waves arriving from all directions are aligned for a given observer. If the observer is displaced by a distance much larger than the wavelength, the propagation phases acquired by waves arriving from different directions are different, and the phases will in general no longer be aligned for the new observer. Thus, a model in which the squeezing phase remains coherent over a large angular region of the sky would single out a particular spatial location. From the viewpoint of the cosmological principle, it is therefore difficult to justify taking the angular size $\Delta\Omega$ of a phase-coherent patch to be large.} The creation and annihilation operators satisfy
\begin{equation}
    \comm{\hat a_i}{\hat a^\dagger_j}=\delta_{ij}\,,
    \quad
    \comm{\hat a_i}{\hat a_j}=0\,.
\end{equation}
Writing $\xi_j=e^{i\phi_j}r_j$, the second-order correlations of each mode are
\begin{equation}
    \braket{\hat a^\dagger_i \hat a_j}=\delta_{ij}\sinh^2 r_j\,,
    \quad
    \braket{\hat a_i \hat a_j}
    =\delta_{ij}e^{i\phi_j}\sinh r_j\cosh r_j\,.
\end{equation}

Because gravitational-wave detectors are not highly directional, the observed gravitational wave is a superposition of waves arriving from many directions.
For example, consider detecting a background gravitational wave arriving uniformly from the entire sky with equal sensitivity. Then, the detector-accessible mode defined in Eq.~\eqref{eq:accecble_mode} is given by
\begin{equation}
    \hat{A} = \sqrt{\frac{\Delta\Omega}{4\pi}}\sum_j\hat a_j\,, 
\end{equation}
which satisfies the normalization condition
\begin{equation}
    \comm{\hat A}{\hat A^\dagger}=1\,. 
\end{equation}
The second-order correlations of the observed operator $\hat A$ are
\begin{equation}
     N_A = \sum_j \frac{\Delta\Omega}{4\pi} \sinh^2 r_j\,,
     \quad
     M_A = \sum_j\frac{\Delta\Omega}{4\pi} e^{i\phi_j}\sinh r_j \cosh r_j\,. 
\end{equation}

Since distinct coherence patches have no common phase reference, their squeezing phases $\phi_j$ are random. In particular, assuming $r_j \sim r$ among the patches and summing all over the entire sky, we have
\begin{align}
    |M_A| \sim \sqrt{\frac{\Delta \Omega}{4 \pi}}\sinh r \cosh r~.
\end{align}
Therefore, the sum over patches reduces the size of $M_A$. Consequently, the witness becomes $W_A\approx-N_A$, which is very large and negative. In other words, squeezing is  washed out in the observed mode.

\subsection{Isolated Squeezed Source}

As we found in the preceding subsection that the phase cancellation among the modes composing the observable mode washes out the squeezing completely, we next consider an even more exotic scenario in which an individual source generates a sufficiently intense single-mode squeezed gravitational-wave state to be observable. As an example of such a source, one may have in mind an astrophysical source such as a binary black-hole merger~\cite{Kanno:2025how,Manikandan:2025dea,Guerreiro:2025mcu} and a superradiant cloud~\cite{Dorlis:2025zzz}. In particular, Ref.~\cite{Manikandan:2025dea} discusses the generation of single-mode squeezing in quasinormal modes. Here, we suppress the polarization label for notational simplicity.

If the gravitational-wave state is represented in a spherical-wave basis centered at the source, it would take the form
\begin{equation}
    \ket{\Psi}=\prod_{\ell m}^{\ellef}
    \expo{\frac12\left(\xi_{\ell m}\hat a^{\dagger2}_{\ell m}- \xi^*_{\ell m}\hat a^2_{\ell m}\right)}\ket{0}\,.
\end{equation}
Because gravitational-wave excitation with high-order multipoles is unlikely, we introduce $\ellef$ as an effective cutoff. The modes with $\ell > \ellef$ are taken to be in their unsqueezed vacuum state. Unless gravitational waves are emitted with extreme directionality comparable to a laser beam, $\ellef$ is expected to be sufficiently small. 

Let us approximate the normalized detector response introduced in Sec.~\ref{sec_linear response} by an angular window function localized within the detector-sensitive region. Consider a sphere centered at the gravitational-wave source with radius $d$, the distance to the observer. Divide this sphere into patches subtending a solid angle $\Delta\Omega$, which corresponds to the area of the detector-sensitive region. Label the patches by $i$, and introduce real-valued window functions $W_i(\Omega)$ normalized by
\begin{equation}
    \int d\Omega\, W_i(\Omega)W_j(\Omega)=\delta_{ij}\,.
    \label{Window_normalization}
\end{equation}
Expanding each window function in spherical harmonics $Y_{\ell m}$ as
\begin{equation}
    W_i(\Omega)=\sum_{\ell,m}c_i^{\ell m}Y_{\ell m}(\Omega)\,,
\end{equation}
the condition in Eq.~\eqref{Window_normalization} implies that the coefficients $c_i^{\ell m}$ satisfy
\begin{equation}
    \sum_{\ell,m}c_i^{\ell m} c_j^{\ell m *}=\delta_{ij}\,.
\end{equation}
The explicit expression for $c_i^{\ell m}$ is
\begin{align}
    c_i^{\ell m} &= \int d\Omega \, Y_{\ell m}^*(\Omega)W_i(\Omega)~.
\end{align}

If the detector lies in the $i$-th patch, the observed mode is expressed using the window function $W_i(\Omega)$ as
\begin{align}
    \hat A &
            = \sum_{\ell,m} \int d\Omega\, W_i(\Omega) \hat a_{\ell m} Y_{\ell m}(\Omega)
            =\sum_{\ell,m} c_i^{\ell m *}\hat a_{\ell m}\,. 
\end{align}
This definition ensures 
$
    \comm{\hat A}{\hat A^\dagger}=1
$.
Since the modes with $\ell>\ell_{\rm eff}$ are in the vacuum state, the second moments of $\hat A$ are
\begin{align}
   & N_A=\sum^{\ellef}_{\ell,m}\abs{c^{\ell m}_i}^2\sinh^2 r_{\ell m}\,,
   \\
   & M_A=\sum^{\ellef}_{\ell,m}\left({c}^{\ell m  *}_i\right)^2  e^{i\phi_{\ell m}} \sinh r_{\ell m}\cosh r_{\ell m}\, .
\end{align}
Therefore, the witness for the effective mode is bounded above as
\begin{equation}
    W_A\leq \sum^{\ellef}_{\ell,m}\abs{c_i^{\ell m}}^2 
    \frac{1-e^{-2r_{\ell m}}}{2}\,.
\end{equation}
From the Cauchy--Schwarz inequality and the addition theorem of spherical harmonics, we have
\begin{align}
	\sum_{\ell,m}^{\ellef} |c_i^{\ell m}|^2 &\le \sum_{\ell,m}^{\ell_{\rm eff}}  \left(\int_{\text{patch $i$}} d\Omega\, |W_i(\Omega)|^2 \right) \left(\int_{\text{patch $i$}} d\Omega\, |Y_{\ell m}(\Omega)|^2\right)\cr
	&= \Delta \Omega \sum_\ell^{\ellef} \frac{2 \ell + 1}{4 \pi}\cr
	&= \Delta \Omega  \frac{(\ellef + 1)^2 - 4}{4 \pi}~.
\end{align}
Here, $-4$ corresponds to the fact that the graviton starts to radiate at $\ell = 2$.
Therefore, we have an upper bound
\begin{equation}
    W_A\leq \Delta \Omega \frac{(\ellef+ 1)^2 - 4}{8\pi}\,.
\end{equation}
Note that the same result holds for spin-weighted spherical harmonics.

When $l_{\rm eff}^2 \Delta \Omega \ll 1$, the witness is suppressed by $\Delta \Omega$. 
Let $L$ be the length scale over which the detector is sensitive. 
Choosing one patch to correspond to the sensitive region of the detector then gives $\Delta\Omega\approx(L/d)^2\ll1$.
For example, the degree of suppression of the witness for a ground-based detector is roughly
\begin{align}
    \Delta \Omega \sim \left(\frac{L}{d}\right)^2 \sim \left(\frac{3 \rm km}{100 \rm Mpc}\right)^2 \sim 10^{-42}\,.
\end{align}
On the other hand, if the source can emit highly collimated gravitational waves with $ \ellef \gtrsim (\Delta \Omega)^{-1/2} \sim 10^{21}$, then the current geometric bound no longer leads to the suppression of the squeezing. Another exotic possibility is generation of squeezed gravitational waves propagating only in one direction as plane waves. However, as far as we are aware, no known mechanism generates such extremely collimated gravitational waves.

\section{Difficulty of Detecting the Quantum Nature of Gravitational Waves}
\label{sec_epsilon}

In the previous section, we found that it is impossible to exploit the squeezing of gravitons arriving from distant sources as an input resource. In this section, we show that the non-classicality of the input gravitons is, in fact, not essential for probing the quantum nature of gravitons.
If the detector is prepared in a controlled squeezed state and deviations from that state can be detected, it is possible to distinguish, through the interaction, whether the graviton is quantum or classical. In doing so, whether or not the graviton itself is squeezed turns out to be largely irrelevant.

\subsection{Quantum and Classical Interaction Models}
Although it is difficult to say that a fully satisfactory model in which gravity is not quantized exists, for the purpose of probing the quantum nature of gravitational waves, one may consider two models: a classical reference model in which the gravitational-wave perturbation is introduced simply as a classical external $c$-number field, and a standard quantum model.
As already discussed in Sec.~\ref{sec_linear response}, in the quantum model the interaction between gravitons and the detector is expected to be given by Eq.~\eqref{eq:quantum_model}, which is rewritten as Eq.~\eqref{Eq:general_interaction} using the creation and annihilation operators of the graviton and the detector.
The corresponding classical interaction model to be contrasted with the quantum model~\eqref{eq:quantum_model} is
\begin{equation}
\hintc = -\frac 12 \int d^3x\, h_{ij}^{\rm TT}\hat T^{ij}\, .
\end{equation}
In terms of the time evolution operator, the classical model is obtained by replacing the operators $\hat A^\dag$ and $\hat A$ in Eq.~\eqref{Eq:general_interaction} by $c$-numbers.

The difference in the linear response of the detector mode between these models affects the witness of the detector mode.
We therefore investigate below the possible values of the detector-mode witness in each model.
As seen in Sec.~2, the presence of a statistical mixture simply reduces the effective squeezing. It is therefore sufficient to consider Gaussian pure states as the initial states of both the gravitational wave and the detector mode.

Our question is whether the final state of the detector can be discriminated from the one reproduced by a classical external field. If the effect of the fundamental graviton--detector interaction can already be described by a classical model, then any subsequent quantum operation performed within the detector cannot rule out a classical origin of the observed state. It is therefore sufficient to examine whether the detector state generated at this elementary interaction stage admits a classical-model description.

\subsection{Detector Witness in the Quantum and Classical Models}

To compare the quantum and classical models, we prepare the detector mode $b$ in a squeezed state. We then evaluate the output witness $W$ in terms of the transformed detector mode $\tilde b$, in which this initial squeezing has been removed. We then examine whether the witness $W_{\tilde b}$\footnote{Note that $W_{\tilde b}>0$ identifies a contraction of the variance along one quadrature relative to the calibrated input covariance, induced by the interaction with gravitational waves, rather than the squeezing initially prepared in the detector.}
can distinguish between the two models even when the gravitational-wave mode itself is not squeezed.

We first consider the quantum model, with the interaction given in Eq.~\eqref{Eq:general_interaction}. As the initial states of the gravitational wave and the detector, we consider squeezed states whose squeezing parameters are respectively given by $\xi_A$ and $\xi_b$:
\begin{equation}
\ket{\Psi_{\rm in}} = \hat S(\xi_A)\ket{0}_A \otimes \hat S(\xi_b)\ket{0}_b\,, 
\end{equation}
where $\ket{0}_A$ and $\ket{0}_b$ are the vacuum states annihilated by $\hat A$ and $\hat b$, respectively. 
We introduce alternative annihilation operators $\hat{\tilde A}$ and $\hat{\tilde b}$, related to the original operators by Bogoliubov transformations, such that
\begin{align}
\hat A & = \hat S^\dagger(\xi_A)\,\hat{\tilde A}\,\hat S(\xi_A)
=\cosh r_A \,\hat{\tilde A}+e^{i\phi_A} \sinh r_A \,\hat{\tilde A}^\dag\,, \cr
\hat b & = \hat S^\dagger(\xi_b)\,\hat{\tilde b}\,\hat S(\xi_b)
=\cosh r_b \,\hat{\tilde b}+e^{i\phi_b} \sinh r_b \,\hat{\tilde b}^\dag\,,
\end{align}
so that the initial state can be regarded as a vacuum state satisfying
\begin{equation}
\hat{\tilde A} \ket{\Psi_{\rm in}}= \hat{\tilde b} \ket{\Psi_{\rm in}} =0\,.
\end{equation}

Expressing the time-evolution operator $\hat{U}_{\rm eff}$ in terms of $\hat{\tilde A}$ and $\hat{\tilde b}$, we obtain
\begin{equation}
	\hat{U}_{\rm eff} = \exp\left(\epsilon\hat A \hat b^\dagger-({\rm h.c.})\right)
= \exp\left(\left(x\hat{\tilde A}+y\hat{\tilde A}^\dagger\right) \hat{\tilde b}^\dagger - ({\rm h.c.})\right)\,,
\end{equation}
with
\begin{align}
&x = \epsilon\left(
\cosh r_A \cosh r_b-e^{i(\phi_b-\phi_A)}\sinh r_A \sinh r_b
\right)\,,
\cr
&y = \epsilon e^{i\phi_A}\left(
\sinh r_A \cosh r_b-e^{i(\phi_b-\phi_A)}\cosh r_A \sinh r_b
\right)\,. 
\end{align}
These coefficients satisfy
\begin{equation}
\abs{x}^2-\abs{y}^2 = \epsilon^2\,.
\label{Eq:x_y_epsilon}
\end{equation}
Using
\begin{equation}
\hat{U}_{\rm eff}^\dagger\, \hat {\tilde b}\, \hat{U}_{\rm eff}
= \cos \epsilon\,\hat{\tilde b}
+ \frac{\sin \epsilon}{\epsilon} \left( x\hat{\tilde A}+y \hat{\tilde A }^\dagger \right)\,,
\end{equation}
the second moments of $\tilde b$ after the interaction are found to be
\begin{equation}
N_{\tilde b} = \left(\frac{\sin \epsilon}{\epsilon}\right)^2\abs{y}^2 , \quad
M_{\tilde b} = \left(\frac{\sin \epsilon}{\epsilon}\right)^2 xy\,,
\end{equation}
and hence the witness is
\begin{equation}
    W_{\tilde b}^{(Q)}
= \left(\frac{\sin \epsilon}{\epsilon}\right)^2 \abs{y}\left(\abs{x}-\abs{y}\right)\,.
\label{Eq:witness_Wq}
\end{equation}
Eq.~\eqref{Eq:x_y_epsilon} implies that $\abs{x}>\abs{y}\geq0$ for a nonzero coupling.
Therefore, the quantum interaction model yields $W_{\tilde b}^{(Q)}\geq0$.
The equality holds only for $\epsilon=0$ or $y=0$, which is the case when $r_b=r_A$ and $\phi_b = \phi_A$.

We next consider the classical model in which the gravitational wave is described by a classical $c$-number field.
As explained above, the evolution operator for the classical case can be obtained by replacing $\hat A$ in $\hat{U}_{\rm eff}$ with a $c$-number $\beta$ as
\begin{equation}
	\hat{U}_{\rm eff}^{\rm (C)} = \exp\left(\epsilon \beta \hat b^\dagger-({\rm h.c.})\right)
= \exp\left(\alpha \hat{\tilde b}^\dagger -({\rm h.c.})\right)\,,
\end{equation}
with $\alpha=\epsilon\left(\beta \cosh r_b-\beta^* e^{i\phi_b}\sinh r_b\right)$.
This is nothing but a displacement operator~\eqref{eq:displace}.
By an argument analogous to that in Sec.~2, when $\alpha$ is simply a constant, the action of the displacement operator does not change the witness, and hence $W_{\tilde b}^{(C)}$ remains zero. When $\alpha$ is instead a classical random variable with a probability distribution, the value of $W_{\tilde b}^{(C)}$ can only decrease. 
Therefore, $W_{\tilde b}>0$ rules out the classical model and thus identifies the quantum interaction model.

The quantum-model witness in Eq.~\eqref{Eq:witness_Wq} can be rewritten as
\begin{equation}
W_{\tilde b}^{(Q)}
= \frac{(\sin \epsilon)^2}{2}\left(1-\left(q-\sqrt{q^2-1}\right)\right)<\frac{(\sin \epsilon)^2}{2}\,,
\label{Eq:witness_Wq2}
\end{equation}
with $q=\cosh 2r_A\,\cosh 2r_b -\cos(\phi_b-\phi_A)\sinh 2r_A\, \sinh 2r_b$. Irrespective of the value of $q$, the upper bound on $W_{\tilde b}^{(Q)}$ is determined by $\epsilon$. Therefore, as long as $\epsilon\ll 1$, the detector witness $W_{\tilde b}^{(Q)}$ cannot be made large. 

Furthermore, even if the gravitational-wave mode $A$ is not squeezed, namely even when $r_A=0$, the value of $q$ can be made large simply by taking $r_b$ sufficiently large, and $W_{\tilde b}^{(Q)}\approx {(\sin \epsilon)^2}/{2}$ can be realized.
Therefore, the squeezing of the gravitational wave itself is not essential for realizing a sufficiently large positive $W_{\tilde b}^{(Q)}$ for the purpose of probing the quantum nature of gravitational waves. We should also notice that, even for a relatively small value of $q$, the quantity in the parentheses gets close to unity, {\it e.g.}, $W_{\tilde b}^{(Q)}\approx 0.366 \times (\sin \epsilon)^2$ for $q=2$.

\subsection{How Strong Must the Coupling be for Detection of the Witness}

In the previous sections, we show that witness is bounded by the coupling strength. Here, we investigate the size of the coupling required for the detection of the witness. Then, we estimate the coupling strength from the proposed gravitational wave detectors and show that it is extremely small for the detection of the witness. 

To demonstrate observationally that $W_{\tilde b}>0$, the value of $W_{\tilde b}$ must be larger than its standard deviation $\Delta W_{\tilde b}$:
\begin{equation}
\label{eq_witness_prove}
W_{\tilde b}\gtrsim \Delta W_{\tilde b}\,.
\end{equation} 
A straight forward computation shows,~\footnote{Notice that the witness $W$ is measuring the (minimum) eigenvalue $\lambda_{-}$ of the covariance matrix $V$. Let $\bm{v}$ be the eigenvector of $\lambda_-$. Then the fluctuation must satisfy
 \begin{align}
    \left(\delta V - \delta \lambda_- I \right)\bm{v} + (V  - \lambda_- I )\delta \bm{v} = 0~,
 \end{align}
 which gives
 \begin{align}
     \delta \lambda_- &= \bm{v} \cdot (\delta V \bm{v})~.
 \end{align}
Thus, $\delta\lambda_-$ is the fluctuation of the quadrature variance along the minimum eigenvalue direction. Note that the quadrature $\Delta X_{\min}$ in the minimum direction follows a Gaussian distribution with zero mean and variance $\lambda_- = 1/2 - W$. Therefore, the variance of the second moment over $N_{\rm trial}$ independent trials is 
\begin{align}
\delta \lambda_-^2 = \frac{1}{N_{\rm trial}}\left[\left\langle(\Delta X_{\min})^4\right\rangle-\left\langle(\Delta X_{\min})^2\right\rangle^2\right]=\frac{2\lambda_-^2}{N_{\rm trial}},
\end{align}
where Wick's theorem, $\langle(\Delta X_{\min})^4\rangle=3\langle(\Delta X_{\min})^2\rangle^2$, has been used. }
\begin{equation}\label{eq:var_witness}
(\Delta W_{\tilde b})^2=\frac{2}{N_{\rm trial}}\left(\frac12- W_{\tilde b}\right)^2.
\end{equation}
Here, $N_{\rm trial}$ is the number of observational trials. 
Then, it follows that, in order to demonstrate $W_{\tilde b}>0$, even when only statistical errors are taken into account, 
\begin{align}
W_{\tilde b}\gtrsim \frac1{\sqrt{2N_{\rm trial}}+2}\,.
\end{align}
Therefore, in combination with Eq.~\eqref{Eq:witness_Wq2}, this implies that
\begin{align}\label{eq:boundeps}
    \epsilon\gtrsim\mathcal{O}(1/N_{\rm trial}^{1/4})\,,
\end{align}
is required in the quantum noise limited case.


\begin{table*}[t]
    \centering
    \caption{
        {
        Representative strain-noise amplitude spectral densities $\sqrt{S_n(f)}$, threshold strains $\hth$, single-graviton strain amplitudes $\hqg$, the noise-normalized coupling strength between the graviton and the detector $\bar{\epsilon}(T_0)$, assuming that the observation duration $T_0$ is set to the graviton oscillation period $\omega_{\rm GW}^{-1}$ for each frequency band. The coupling $\bar{\epsilon}(T_0)$ is derived from $\bar \epsilon(T_0)=h_{\rm QG}(T_0)/h_{\rm th}(T_0)$.   
        The quoted values of $\sqrt{S_n(f)}$ correspond to the best sensitivities of selected planned detectors in the respective frequency bands.
        }
    }
    \label{tab: sensitivity}

    \renewcommand{\arraystretch}{1.3}
    \setlength{\tabcolsep}{8pt}

    \begin{tabular}{c | c | c | c | c}
        \hline\hline
        $\omegagw(=T_0^{-1})$
        & $\sqrt{S_n(f)}\,[\mathrm{Hz}^{-1/2}]$ & $h_{\rm th}(T_0)$ & $h_{\rm QG}(T_0)$ & 
        $\bar \epsilon(T_0)$
        \\ \hline\hline
        
        $\mathrm{10\, nHz}$
        & $10^{-14}$\,\cite{Janssen:2014dka} & $10^{-18}$ & $10^{-51.5}$ & $10^{-33.5}$
        \\

        $1\, \mathrm{mHz}$
        & $10^{-20}$\,\cite{Robson:2018ifk} & $10^{-21.5}$ & $10^{-46.5}$ & $10^{-25}$
        \\

        $0.1$--$1\,\mathrm{Hz}$
        & $10^{-24}$\,\cite{Harry:2006fi} & $10^{-24}$ & $10^{-43.5}$ & $10^{-19.5}$
        \\

        $20$--$400\,\mathrm{Hz}$
        & $10^{-25}$\, \cite{Evans:2021gyd} & $10^{-24}$ & $10^{-41.5}$ & $10^{-17.5}$
        \\

        $1\,\mathrm{kHz}$
        & $10^{-25}$\,\cite{Srivastava:2022slt} & $10^{-23.5}$ & $10^{-40.5}$ & $10^{-17}$
        \\

        $100\,\mathrm{kHz}$
        & $10^{-22}$\,\cite{Aggarwal:2020umq} & $10^{-19.5}$ & $10^{-38.5}$ & $10^{-19}$
        \\

        $100\mathrm{MHz}$
        & $10^{-20}$\,\cite{Nishizawa:2007tn} & $10^{-16}$ & $10^{-35.5}$ & $10^{-19.5}$
        \\

        $\mathrm{10GHz}$
        & $10^{-21}$\,\cite{Domcke:2024eti}
        & $10^{-16}$ & $10^{-33.5}$ & $10^{-17.5}$
        \\

        $1\, \mathrm{THz}$
        & $10^{-19.5}$\,\cite{Kahn:2023mrj} 
        & $10^{-13.5}$ & $10^{-31.5}$ & $10^{-18}$
        \\

        \hline\hline
    \end{tabular}
\end{table*}

To assess whether the graviton--detector coupling $\epsilon$ can be large enough to satisfy the requirement ~\eqref{eq:boundeps}, we consider a wave packet mode of a gravitational wave with angular frequency $\omega_{\rm GW}$, cross-section ${\cal A}$, and duration $T$. 
For such a normalized wave-packet mode $A$, the corresponding strain-amplitude operator can be written as
\begin{equation}
\hat{h}(T) = h_{\rm QG}(T)\hat{X}_A\, ,
\end{equation}
where 
\begin{align}
	h_{\rm QG}(T) \sim \frac{1}{M_{\rm pl} \sqrt{\omega_{\rm GW} {\cal A} T}}\,.
\end{align}
 Here, $\hat{X}_A$ is the dimensionless quadrature operator of the graviton. Note that wave packet mode function $u_A$ must be normalized such that $\int d^3 x\, u_A^2 \sim M_{pl}^{-2}\omega_{\rm GW}^{-1}$, and the volume of the wave packet is identified as $\mathcal{A} T$. For the estimates below, we take the cross-section of the wave packet to match the wavelength of the gravitational wave ${\cal A}\sim\omegagw^{-2}$, as a possible minimum value.

To leading order in the coupling, the incident wave changes the quadrature of the detector $\hat{X}_b$ through the interaction in Eq.~\eqref{Eq:general_interaction},
\begin{align}
\delta\braket{\hat X_b}
&=\braket{\hat X_{b,{\rm out}}} - \braket{\hat X_{b,{\rm in}}}
\sim \epsilon\braket{\hat X_A}
=\epsilon\frac{h(T)}{\hqg(T)}\,.
\label{Eq:deltax_b}
\end{align}
Let $\sigma_b^2=\langle(\Delta\hat X_b)^2\rangle$ denote the variance of the normalized detector output. 
The condition for the detectability of the change is 
\begin{align}\label{Eq:thres_Xb}
\abs{\delta\langle\hat X_b\rangle}\gtrsim\sigma_b\,.
\end{align}
We define $\hth(T)$ as the threshold strain amplitude that saturates this condition. Combining conditions.~\eqref{Eq:deltax_b} and \eqref{Eq:thres_Xb}, we then define the noise-normalized coupling strength as
\begin{align}
\bar{\epsilon} \coloneqq \frac{\epsilon}{\sigma_b}\simeq\frac{\hqg(T)}{\hth(T)}\,. 
\label{Eq:def_barepsilon}
\end{align}
Here, the same threshold strain can be expressed in terms of the noise spectral density $S_n$ by requiring the signal-to-noise ratio from matched filtering
\begin{align}
\rho^2=4\int_0^\infty df\,\frac{\abs{h(f)}^2}{S_n(f)}\, ,
\label{Eq:snr}
\end{align}
to be of order unity. When we set the observation duration to the gravitational wave period, {\it i.e.}, $T_0=\omega_{\rm GW}^{-1}$, the threshold becomes 
\begin{align}
	h_{\rm th}(T_0) \sim\sqrt{\omegagw S_n(f)}~.
\label{Eq:strain_sens}
\end{align}
Thus, Eqs.~\eqref{Eq:def_barepsilon} and \eqref{Eq:strain_sens} show that a strain sensitivity determines the noise-normalized coupling $\bar{\epsilon}$, rather than the microscopic mixing strength $\epsilon$ itself. In Table~\ref{tab: sensitivity}, we show the estimates of $\bar{\epsilon}$ at the reference duration $T_0=\omegagw^{-1}$ for each frequency, based on the sensitivity of proposed detectors. The values in Table~\ref{tab: sensitivity} therefore imply $\bar\epsilon(T_0)\ll1$ throughout the frequency range considered.

To obtain the bound on the coupling $\epsilon$ from the estimates of $\bar \epsilon$, we consider the most favorable case in which the detection is quantum-noise limited. Any additional noise would increase the uncertainty of the witness $W_{\tilde b}$, and hence removal of additional noise is inevitable. In the following, we assume such noise is below the quantum-noise level $(\sigma_b=O(1))$, so that $\bar{\epsilon}$ provides an order-of-magnitude estimate of $\epsilon$. We henceforth do not distinguish between $\bar{\epsilon}$ and $\epsilon$.

For completeness, we consider a more optimistic scenario, in which the signal caused by the interaction with the gravitational wave mode can be accumulated coherently for a longer time $T \gg T_0$. In such a case, the change in the detector quadrature increases with time as $\delta \braket{X_b}\propto hT$, as long as the detector remains quantum-noise limited.
Then, the threshold decreases as $h_{\rm th} \propto T^{-1}$, while the single graviton amplitude only decreases as $h_{\rm QG} \propto T^{-1/2}$, and thus the coupling increases to
\begin{align}
    \epsilon(T)\sim\sqrt{\omegagw T}\,\epsilon(T_0)\,.
\label{Eq: epsilon_t}
\end{align}
Even in this optimistic scenario, the gain is not so significant. A THz-band signal coherently observed for one year gives only a factor $10^{10}$, and hence $\epsilon(1\,{\rm yr})\sim10^{-8.5}$ for the representative values in Table~\ref{tab: sensitivity}. 

Finally, let us show that dividing a fixed total observing time $T_{\rm tot}$ into $N_{\rm trial}=T_{\rm tot}/T_{\rm obs}$ statistically independent intervals does not improve the detectability of the witness. Increasing $N_{\rm trial}$ improves the statistical precision. According to Eq.~\eqref{eq:boundeps}, the required coupling for each interval decreases as $\epsilon(T_{\rm obs})\gtrsim N_{\rm trial}^{-1/4}$.
On the other hand, under the optimistic scaling in Eq.~\eqref{Eq: epsilon_t}, the coupling accumulated within each interval decreases as $\epsilon(T_{\rm obs})=\epsilon(T_{\rm tot})/\sqrt{N_{\rm trial}}$. Combining these scalings, we obtain
\begin{align}
    \epsilon(T_{\rm tot})\gtrsim N_{\rm trial}^{1/4} \gtrsim 1\,.
\end{align}
Thus, subdividing a fixed observing time into many independent modes does not compensate for weak conversion. Rather, a small number of coherent samplings $N_{\rm trial} \sim 1$ is preferred, if possible.

\section{Conclusion}
\label{sec_conclusion}

In this paper, we investigated two challenges in the detection of the quantum nature of gravitational waves, characterized by the squeezing witness. The first is the dilution of the quantum witness when a state prepared at the source is projected onto the wave packet mode accessible by a detector. The second arises when considering the transfer of a quantum signature to the detector mode through the weak gravitational coupling. Throughout the study, we restricted our consideration to Gaussian quantum states and a linear coupling to a single detector mode, which we believe gives the most optimistic estimate. Furthermore, in the classical model, to discriminate from the quantum model, the gravitational wave was treated as a $c$-number field.

First, even when the global gravitational wave state is highly squeezed at the source, this squeezing is not generally retained by the effective mode coupled to a local detector.  For inflationary gravitational waves, the detector accesses only one part of each two-mode squeezed pair and therefore observes a thermal marginal state. For a squeezed stochastic background, the superposition of modes with random squeezing phases washes out the observable squeezing.  Even if we assume an isolated source of a squeezed gravitational wave, the finite angular region accessible by the detector strongly suppresses the upper bound on the witness.  These examples show that source-level squeezing is not by itself an observable resource; it must be evaluated after projection onto the actual detector-coupled mode.

Second, squeezing of the incident gravitational wave is not necessary in principle to distinguish a quantized field from a classical $c$-number drive. If the detector is prepared in a controlled squeezed state, its response to a quantized gravitational-wave mode can acquire a positive witness even when the incident mode is initially unsqueezed or even unexcited.  By contrast, a prescribed classical field only displaces the detector state and cannot generate such a positive witness.  The quantum response is nevertheless bounded by $W_{\tilde b}<\sin^2\epsilon/2\sim \epsilon^2/2$.  Detector squeezing can
therefore remove the need for a non-classical incident state, but it cannot remove the parametric suppression due to the weak gravitational coupling. The optimistic sensitivity estimates considered here give $\bar{\epsilon}\ll1$ over the full frequency range, and neither coherent integration nor a subdivision of a fixed observing time into independent samples overcomes this suppression within the present setup.

Taken together, these results clarify the relation between the two difficulties.  A strategy that attempts to observe non-classical radiation produced at the source faces both the reduction to an accessible mode and the weak transfer to the detector.  Preparing the detector in a non-classical state can bypass the first difficulty, but not the second.  The relevant experimental target is therefore not stronger squeezed gravitational-wave sources, but quantum-coherent transduction capable of resolving the effect of the vacuum fluctuation of the gravitational-wave mode coupled to the detector.  In this sense, detecting a gravitational-wave signal and detecting the quantum nature of the gravitational-wave field require different sensitivity criteria.

\section*{Acknowledgements}
We thank Shogo Tomizuka for valuable discussions on the detectability of gravitational-wave squeezing with laser interferometers, detector-accessible effective wave-packet modes, and the interaction between gravitational waves and detectors. Two of the authors (Y. M. and H. T. ) also acknowledge discussions with him in the context of a separate collaboration, during which a calculation note containing related results was shared.
Y.~M.~is supported by JST SPRING, Grant Number JPMJSP2110. 
H.~O.~is supported by JSPS KAKENHI Grant Numbers JP23H00110 and JP25K17388. 
A.~O.~is supported in part by the National Natural Science Foundation of China under Grant No. 12403001 and 12547101, and New Chongqing YC Project CSTB2024YCJH-KYXM0083.
H.~T.~is supported by the Hakubi project at Kyoto University, and also by JSPS KAKENHI Grant No. JP26K17146.
T.~T.~is supported by Grant-in-Aid for Scientific Research under Contract No.~JP23H00110, and also by SPIRIT2 2026 of Kyoto University. 

\begingroup
\normalem
\bibliographystyle{JHEP}
\bibliography{sankou}
\endgroup

\end{document}